\documentclass[aps,onecolumn,prd,nofootinbib]{revtex4}
\usepackage{amsmath}
\usepackage{graphicx}
\usepackage{dcolumn}
\usepackage{bm}
\usepackage{amssymb}
\usepackage{latexsym}

\begin{document}

\title{Gravitational Higgs mechanism in inspiraling scalarized NS-WD binary}

\author{J. Wang\footnote{Email address: wangjing6@mail.sysu.edu.cn}}

\affiliation{School of Physics and Astronomy, Sun Yat-Sen University,
   Guangzhou, 510275, P. R. China}

\begin{abstract}
We investigate the gravitational Higgs mechanism in the inspiraling scalarized neutron star - white dwarf (NS-WD) binaries, whose dynamics are described by the scalar-tensor theory. Because of the difference in binding energy of NS and WD, the orbital decay of scalarized NS-WD system actually sources an emission of dipolar gravitational scalar radiation, in addition to the tensor gravitational waves, which breaks the Lorentz invariance constructed in the framework of general relativity. The resulted gravitational scalar radiation field obtains a scalar-energy-density-dependent effective mass, arising from a gravitational scalar potential that consists of a monotonically decreasing self-interactions of gravitational scalar field and an increasing exponential coupling between the scalar field and the NS/WD matter. Owing to a thin-ring-orbit effect, the gravitational interactions encoded by the massive scalar field is screened in the region of binary orbit, with high density of stars' scalar energy, which gives us the estimation for scalar masses of about $10^{-21} eV/c^2$ and leads to a Yulkawa-like correction to the Newtonian potential of the binary system. We demonstrate that the radiated gravitational tensor waves, propagating in the Yukawa type of potential, gain a scalar-background-dependent mass term of the order of $\sim 10^{-23} eV/c^2$.  
\end{abstract}


\maketitle


\section{Introduction}

Neutron star - white dwarf (NS-WD) binaries usually contain a massive recycled NS \cite{2011A&A...527A..83Z, 2010Natur.467.1081D, 2013Sci...340..448A}, which results from the recycling process \cite{2011A&A...526A..88W}. A massive NS tends to undergo spontaneous scalarization,  which is analogous to the ferromagnetism \cite{Damour:1993hw, Damour:1996ke}. Accordingly, a nontrivial scalar configuration settles inside the NS and acts as a source for gravitational field, which contributes to a gravitational scalar interaction with its companion and induces a scalarization inside the WD \cite{Palenzuela:2013hsa}.  Consequently, external scalar fields $\varphi_{\it NS}$ and $\varphi_{\it WD}$ appear around the scalarized NS and WD, respectively. The scalar configurations lead to modifications of the microphysics in the interiors of relativistic stars, and the binding energy, as a significant portion of the gravitational mass of the stars, is subsequently changed. The gravitational interaction between the spontaneous scalarized NS and the induced scalarized WD is enhanced, and the Newtonian gravitational interaction of the binary is modified according to \cite{Damour:1996ke}
\begin{equation}
V_{\it int} = -G\frac{m_{\it NS}m_{\it WD}}{r_{\it NS-WD}} - G\frac{\omega_{\it NS}\omega_{\it WD}}{r_{\it NS-WD}},
\end{equation}
where $m_{\it NS}$ and $m_{\it WD}$ represent the masses of the NS and the WD, $\omega_{\it NS}$ and $\omega_{\it WD}$ denote the scalar charges of corresponding components with the definition of $\omega_i = - \frac{\partial \ln m_i(\varphi_i)}{\partial \varphi_i}$\footnote{Note that we use $i$ on the subscript or index to denote NS or WD in the binary system all over the paper.} \cite{Damour:1992we}, $G$ is the Newtonian gravitational constant, and $r_{\it NS-WD}$ is the orbital separation of the binary. The local Newtonian gravitational constant is accordingly modified as
\begin{equation}
G_{\it eff} = G (1 + \omega_{\it NS}\omega_{\it WD} + \cdots),\label{effG}
\end{equation}
which is assigned to be the effective gravitational constant of a scalarized NS-WD binary system. The second term in the bracket, $\omega_{\it NS}\omega_{\it WD}$, is the post-Newtonian corrections to the Newtonian binary system, and the "$\cdots$" denotes the terms of dissipative corrections to the Newtonian dynamics that accounts for the backreaction of gravitational-wave emission. 

It is well known that NS is more compact than WD. Therefore, the dependence of star masses on the scalar fields are different, which actually sources an emission of dipolar gravitational scalar radiation in a post-Newtonian inspiraling scalarized NS-WD binary \cite{Palenzuela:2013hsa}, with an energy flux $\dot{E}^{\it dipole} \propto (\omega_{\it NS} - \omega_{\it WD})^2$. We assign the associated radiated scalar particles to be the scalarons. Observationally, the long baseline of precise timing observations for PSR J1738+0333 \cite{2012MNRAS.423.3328F} have indicated an excess orbital decay of $+2.0^{+3.7}_{-3.6}\it fs~s^{-1}$, which directly translates to a constraint on deviations from the quadruple formula.  The Lunar Laser Ranging experiments give the limit on dipole radiation of $\dot{P}^{\it dipole} = 1.9^{+3.8}_{-3.7}\it fs~s^{-1}$.

As a result, the dynamics of a scalarized NS-WD system is governed by a gravitational scalar radiation field $\phi$, together with the gravitational tensor metric $g_{\mu\nu}$. The scalar charge of a scalarized NS-WD binary can be extracted from the behavior of the gravitational scalar field near spatial infinity \cite{Damour:1992we}, i.e.
\begin{equation}
\phi = \phi_0 +\frac{\phi_1}{r} + \mathcal{O}(\frac{1}{r^2}),\label{sbehavior}
\end{equation}
where we consider the iterative gravitational interplay between secularized NS and WD and a convergence of the external scalar fields $\varphi_{\it NS}$ and $\varphi_{\it WD}$, and $\phi_0$ is the asymptotic value of the gravitational scalar field. We use $r$ to denote the distance from the center of the NS-WD binary. 

As we know that the radiation of quadruple gravitational waves is a representative of the Lorentz invariance in a relativistic system described by general relativity. In a scalarized inspiraling NS-WD binary, with both dipolar gravitational scalar radiation and the quadruple gravitaional tensor radiation, the Lorentz invariance of the relativistic binary is spontaneously broken. According to the Higgs mechanism, a spontaneous symmetry breaking is essential to a generation mechanism of property "mass" for associated particles. In this work, we investigate the gravitational Higgs mechanism in post-Newtonian inspiraling scalarized NS-WD systems and demonstrate the mass generation mechanism for associated particles resulting from the Lorentz symmetry breaking. In section \ref{sec:action}, we describe the dynamics of the scalarized NS-WD binaries in the framework of scalar-tensor theory. Section \ref{sec:scalmass} discusses the mass generation of scalarons and the effects of massive gravitational scalar field. In order to study the influence of massive scalarons on tensor gravitational waves, we consider the solutions of the gravitational scalar field outside the binary systems in section \ref{sec:scalsolu}. The mechanism of scalar-mediated mass origin for gravitons is demonstrated in section \ref{sec:gravmass}. Finally, we give some remarks and discussions in section \ref{sec:rd} for the gravitational Higgs mechanism in scalarized NS-WD binaries and effects in specific systems involving more massive NSs .

\section{The action}\label{sec:action}

The dynamics of an inspiraling scalarized NS-WD binary system, suffering from the post-Newtonian corrections, is described by a scalar-tensor theory with the action
\begin{eqnarray}
S = &&\int d^4x \sqrt{-g} (\frac{\mathcal{R}}{16\pi G} - \frac{1}{2}g^{\mu\nu}\partial_{\mu}\phi\partial_{\nu}\phi - V(\phi))\nonumber\\ &&+ \sum\limits_n\int_{\gamma_n} ds~m_{i_n}(\phi).\label{action} 
\end{eqnarray}
Here, $\mathcal{R}$ and $g$ are the Ricci scalar and the determinant of the metric $g_{\mu\nu}$, respectively. 

The scalar potential $V(\phi)$ results from two interactions, i.e. the self-interactions of $\phi$ and the interactions between $\phi$ and matter fields of NS/WD. The gravitational scalar field is associated with the non-perturbative strong-field effects \cite{Damour:1993hw}, which contributes to a potential of the runaway form \cite{Khoury:2003rn} that satisfies $\lim\limits_{\phi\rightarrow\infty} V(\phi) \to 0,~\lim\limits_{\phi\rightarrow\infty} \frac{V(\phi)'}{V(\phi)} \to 0,~\lim\limits_{\phi\rightarrow\infty} \frac{V(\phi)''}{V(\phi)'} \to 0,~...$, as well as $\lim\limits_{\phi\rightarrow 0} V(\phi) \to \infty,~\lim\limits_{\phi\rightarrow 0} \frac{V(\phi)'}{V(\phi)} \to \infty,~\lim\limits_{\phi\rightarrow 0} \frac{V(\phi)''}{V(\phi)'} \to \infty,~...$ ($V(\phi)' \equiv \frac{dV}{d\phi}$, and $V(\phi)'' \equiv \frac{d^2V}{d\phi^2}$, etc.). Thus, the self-interactions of gravitational scalar field, whose  behavior is described by Eq. (\ref{sbehavior}), lead to a monotonically decreasing potential,
\begin{equation}
V_{\phi} = \mu^5/\phi,\label{dpotential}
\end{equation}
where $\mu$ has the unit of mass. The NS/WD matter interacts directly with the gravitational scalar field $\phi$ through a conformal coupling of the form $e^{-\alpha_i\phi/M_{pl}}$. Here, $M_{pl} \equiv (8\pi G)^{-1/2}$ is the reduced Planck mass. The dimensionless coupling constant $\alpha_i \equiv -M_{pl}\frac{d\log m_i}{d\phi}$ describes the measure of the coupling strength between star matter and the scalar field, whose values are usually negative in observations \cite{Damour:1990tw}. So the exponential coupling function is an increasing function of $\phi$. The combined effects of self-interactions of $\phi$ described by Eq. (\ref{dpotential}) and the conformal coupling give us the form of the scalar potential $V(\phi)$ in Eq. (\ref{action}),
\begin{equation}
V(\phi) = \frac{\mu^5}{\phi} + \varepsilon_{\varphi_i}e^{-\alpha_i\phi/M_{pl}},\label{potential}
\end{equation}
where $V(\phi)$ is an explicit function of energy density of the stars' scalar configurations $\varepsilon_{\varphi_i}$. The reason is that the coupling between gravitational scalar field and star matter can lead to modifications of the microphysics in the interiors of NS and WD, which results in fluctuations of external scalar fields $\varphi_i$ and gradient of scalar energy density during the orbital motion. The changes of strength for interior scalar configurations and the condensation effects in the binary orbit subsequently have influence on the gravitational scalar potential. $\varepsilon_{\varphi_i}$ depends on the masses of the stars (a function of the density for each star $\rho_i$) and the coupling strength between scalar configuration and components in the interior of NS/WD \cite{Damour:1996ke}.

The second line of Eq. (\ref{action}) describes the action of matter components making up the NS and the WD. In the sum over $n$ we give the world line action for the number of any species of matter and particles consisting in the NS and the WD and we use $\gamma_n$ to represent the integral of the matter action along world line. The couplings of matter components inside the stars to the scalar field arise from the dependence of the masses $m_i$ on $\phi$. The NS/WD matter couples to the gravitational tensor metric $g_{\mu\nu}$ via the conformal transformation $e^{-\alpha_i\phi/M_{pl}}$, according to the rescaling relation,
\begin{equation}
g^{*}_{\mu\nu} = e^{-2\alpha_i\phi/M_{pl}}g_{\mu\nu}.
\end{equation} 

\section{The mass of scalarons}\label{sec:scalmass}

The combined scalar potential $V(\phi)$ expressed in Eq. (\ref{potential}) consists of a monotonically decreasing potential (\ref{dpotential}) and a monotonically increasing interaction $e^{-\alpha_i\phi/M_{pl}}$, which actually displays a minimum. By minimizing the differentiation of the scalar potential with respect to $\phi$, i.e.
\begin{equation}
V(\phi)' - \sum\limits_i\frac{\alpha_i}{M_{pl}}\varepsilon_{\varphi_i}e^{-\alpha_i\phi/M_{pl}} =0,\label{phimin}
\end{equation}
we can get the value of $\phi$ at the minimum potential $\phi_{\min}$. Around this minimum, the gravitational scalar field acquires an effective mass, which is obtained by evaluating the second derivative of the potential at $\phi_{\min}$,
\begin{equation}
m_s^2 = V(\phi)''|_{\phi_{\min}} + \sum\limits_i\frac{\alpha_i^2}{M_{pl}^2}\varepsilon_{\varphi_i}e^{-\alpha_i\phi/M_{pl}}|_{\phi_{\min}}.\label{smass}
\end{equation}
Equations (\ref{phimin}) and (\ref{smass}) imply that both the local value of  gravitational scalar radiation field $\phi_{\min}$ and the mass of scalarons depend on the local energy density of the stars' scalar configurations. It can be found, from Eq. (\ref{smass}), that the scalarons become more massive in a higher $\varepsilon_{\varphi_i}$ environment. Therefore, the mass mode of the gravitational scalar field propagates effectively when the ambient scalar environment displays a higher energy density.

Most of the NS-WD binaries have very small orbital eccentricity of $\sim 10^{-5} - 10^{-6}$ \cite{2011A&A...527A..83Z}, i.e. approximate circular orbits. Accordingly, the scalarized NS and WD orbit with each other and form an orbit ring on the binary plane, with depositing scalar energy. The scalar charges of the system is given by $Q = -\frac{\ln\partial M_c(\varphi)}{\partial\varphi}$, where $M_c(\varphi)$ is the reduced mass of the binary and is modified according to the couplings between NS/WD matter fields and $\phi$ \cite{Salgado:1998sg}, $M_c = \frac{m_{\it NS}(\varphi)~m_{\it WD}(\varphi)}{m_{\it NS}(\varphi) + m_{\it WD}(\varphi)}$. We then assume the orbiting binary system as a ring configuration with semi-spherically symmetric boundary. The distance $a$ from the center of the binary plane to the outer boundary of the ring configuration corresponds to the semi-separation of an NS-WD binary, which is of the order of $\sim 10^9$ m \cite{2011A&A...527A..83Z}, and the central thickness of the orbital ring approximately equals the diameter of the WD, i.e. $\Delta a \sim 2 R_{\it WD} \sim 10^6$ m. The motion of scalarized stars yields a deposited scalar energy of $E_{\varphi} = M_c(\varphi) + \frac{1}{2}cQ^2$ \cite{Damour:1996ke} in the region of orbital ring. The coefficient $c$ is related to the mass and scale of the binary, as well as the coupling strength to the gravitational scalar field, $c = \frac{1}{a}(1+\sum\limits_i\alpha_i\frac{M_c(\varphi)}{a})$. By comparing the radius of NS and WD with the separation between them, we can get $\frac{\Delta a}{a} \sim 10^{-3} \ll 1$. Accordingly, the orbit of NS-WD binary can be assigned to be a thin-ring orbit.

The gravitational interaction between scalarized NS and WD, encoded by a massive gravitational scalar field, typically acquires an exponential Yukawa suppression, which results in a finite range of Yukawa type of potential energy,
\begin{equation}
U(r) = -2\alpha_{\it NS}\alpha_{\it WD}\frac{Gm_{\it NS}m_{\it WD}}{r}e^{-m_sr},\label{YukPot}
\end{equation}
where the product $2\alpha_{\it NS}\alpha_{\it WD}$ is referred to as the interaction strength. The inverse of the interaction range $\lambda$ for a Yukawa potential $e^{-r/\lambda}/r$ characterizes the mass of scalarons $\lambda^{-1} \equiv m_s$. 
As a consequence of the thin-ring orbit, the gravitational scalar interaction between NS and WD is screened in an interaction range of the same order of the orbital width $\lambda \sim 2R_{WD} \sim10^6$ m. Accordingly, the corresponding mass of scalarons is estimated as $m_s \equiv \lambda^{-1} \sim \Delta a^{-1} \sim 10^{-21} eV/c^2$. The ratio $\frac{\Delta a}{a}$ is assigned to be the screening parameter, which can be understood as playing the role in screening the propagations of mass mode of the massive gravitational scalar field. We refer to such a screening mechanism in NS-WD binaries as the thin-ring-orbit effect.

\section{Solutions of the gravitational scalar field}\label{sec:scalsolu}

The equation of motion (e.o.m.) for the gravitational scalar field can be derived by varying the action (\ref{action}) with respect to $\phi$, which reads
\begin{equation}
\Box\phi = V'_{\phi} - \sum\limits_i\frac{\alpha_i}{M_{pl}}\varepsilon_{\varphi_i}e^{-\alpha_i\phi/M_{pl}},\label{seom}
\end{equation} 
where $\Box = g^{\mu\nu}\partial_{\mu}\partial_{\nu}$ is the d'Alembert's operator. 

In the framework of thin-orbit-ring configuration for scalarized NS-WD binary system, with spherically symmetric semi-sphere outer boundary, we consider an infinitesimal volume elements within the orbital ring. The gravitational scalar field can be solved in a static, spherically symmetric regime. Accordingly, the e.o.m. (\ref{seom}) is reduced to
\begin{equation}
\frac{d^2\phi}{dr'^2} + \frac{1}{r'}\frac{d\phi}{dr'} = V'_{\phi} + \sum\limits_i\frac{\alpha_i}{M_{pl}}\varepsilon_{\varphi_i}e^{\alpha_i\phi/M_{pl}},\label{ssseom}
\end{equation}
where the coordinates $r'$ denotes the distance from the center of orbital ring for the binary. This differential equation (\ref{ssseom}) is subject to the following boundary conditions,
\begin{eqnarray}
\phi &=& \phi_{out},~~~\it as~r' = \frac{\Delta a}{2};\nonumber\\\label{condition}
\phi &=& \phi_{in},~~~\it as~0 \leq r' \leq \frac{\Delta a}{2};\nonumber\\
\phi &=& \phi_0,~~~\it as~r' \rightarrow \infty.
\end{eqnarray}
Here, the thickness of the orbital ring $\Delta a$ is related to $\phi_{out}$, $\phi_{in}$, and the Newtonian potential of the binary system $\Phi_c = \frac{M_c}{8\pi M_{pl}^2a}$, which is given by \cite{Khoury:2003rn}
\begin{equation}
\frac{\Delta a}{a} = \frac{\phi_{out} - \phi_{in}}{6\sum\limits_i\alpha_iM_{pl}\Phi_c}.\label{thick} 
\end{equation}

In analogy with the electrostatic shield of an electronic conducting shell, the deposited scalar energy is screened and dominates in the ring orbit, and the scalar charges are distributed on the surface of outer boundary with the radius of $r'=\frac{\Delta a}{2}$. Therefore, the gravitational scalar field inside the ring $\phi_{in}$ can be considered as perturbations, i.e. $\phi_{in} \ll \phi_{out}$. 

Aiming to investigate the influence of the massive gravitational scalar field on the propagations of gravitons, we are just interested in the solutions outside the system, i.e. solutions in the region of $r' > \frac{\Delta a}{2}$. By solving Eq. (\ref{ssseom}) and using the boundary conditions (\ref{condition}), we get the exact exterior solutions,
\begin{equation}
\phi(r'>\frac{\Delta a}{2}) = \phi_{out}(1 - \frac{\phi_{out} - \phi_{in}}{6\sum\limits_i\alpha_iM_{pl}\Phi_c})\frac{ae^{-m_s(r'-\frac{\Delta a}{2})}}{r'} + \phi_{out}.\label{solution}
\end{equation}
By considering the field density contrast $\phi_{in} \ll \phi_{out}$ and in the limit of thin-ring orbit $\Delta a \ll a$, the combination of Eqs. (\ref{thick}) and (\ref{solution}) gives the approximative solutions
\begin{equation}
\phi(r'>\frac{\Delta a}{2}) \approx \frac{\sum\limits_i\alpha_i}{4\pi M_{pl}}\frac{3\Delta a}{a}\frac{M_ce^{-m_sr'}}{r'} + \phi_{out}.\label{asolution}
\end{equation}

\section{Scalar-background-dependent mass of gravitons}\label{sec:gravmass} 

Variation of the action (\ref{action}) with respect to the metric gives us the following e.o.m.,
\begin{eqnarray}
(\mathcal{R}_{\mu\nu}-\frac{1}{2}g_{\mu\nu}\mathcal{R})M_{pl}^2 = \partial_{\mu}\phi\partial_{\nu}\phi-\frac{1}{2}g_{\mu\nu}(\partial_{\alpha}\phi)^2+g_{\mu\nu}(\frac{\mu^5}{\phi} + \varepsilon_{\varphi_i}e^{-\alpha_i\phi/M_{pl}}).\label{geom}
\end{eqnarray}
We consider the weak-field scalar and tensor perturbations, i.e. $g_{\mu\nu} = \eta_{\mu\nu} + h_{\mu\nu}$, $\phi = \phi_{out} + \delta\phi$, as well as the small perturbative coupling $\sigma^{\mu\nu} = h^{\mu\nu} - \frac{1}{2}h\eta^{\mu\nu} - \frac{\delta\phi}{\phi_{\it out}}\eta^{\mu\nu}$. Expanding the left-hand side of Eq. (\ref{geom}) in the weak field limits, we rewrite the e.o.m. of gravitons as 
\begin{eqnarray}
\Box_{\eta}\bar{h}_{\mu\nu} + \frac{1}{2}\Box_{\eta}\sigma_{\mu\nu} + \eta_{\mu\nu}\Box_{\eta}(\frac{\delta\phi}{\phi_{\it out}}) = \frac{1}{M_{pl}^2}(\partial_{\mu}\delta\phi\partial_{\nu}\delta\phi - \frac{1}{2}\eta_{\mu\nu}(\partial_{\alpha}\delta\phi)^2) + \frac{1}{2M_{pl}^2}\eta_{\mu\nu}m_s^2(\phi-\phi_{out})^2,\label{feom}
\end{eqnarray}
where we impose the harmonic gauge conditions of $\partial^{\nu}(h_{\mu\nu} - \frac{1}{2}\eta_{\mu\nu}h) = 0$ and $\partial^{\nu}{\sigma_{\mu\nu}} = 0$, and the expansion of $V(\phi)$ in Taylor series about $\phi_{out}$ is also used. $\Box_{\eta} = \eta^{\mu\nu}\partial_{\mu}\partial_{\nu}$ is the flat-space D'Alembertian, $h = \eta^{\mu\nu}h_{\mu\nu}$, and $\bar{h}_{\mu\nu} = h_{\mu\nu} - \frac{1}{2}\eta_{\mu\nu}h$. 

We then substitute the exterior approximative solutions (\ref{asolution}) of $\phi$ into Eq. (\ref{feom}) and follow the gauge selection described in \cite{Maggiore2008}. It is evident that the motion of gravitons has the wave solutions, with the modifications resulting from the gravitational scalar radiation \cite{Will1993, Alsing:2011er},
\begin{eqnarray}
\bar{h}_{\mu\nu} = \int d\omega\int\frac{d^3\vec{k}}{(2\pi)^3}Ae^{i(\vec{k}\cdot\vec{r}-\omega t)}\cdot\int d\omega'e^{i(k'_rr-\omega' t)} \phi_{out}(\omega')\frac{M_c}{r}(1+\sum\limits_i\alpha_ie^{-m_sr}).\label{gws}
\end{eqnarray}
Here, we consider the plane-wave solutions of the gravitational scalar radiation $\phi = \int\phi_{out}(\omega')e^{i(k'_rr-\omega' t)}d\omega'$ \cite{Wagoner:1970vr}. The quantities $\omega$, $\vec{k}$, and $A$ denote the frequency, the wave vector and the amplitude of tensor gravitational waves radiated from the orbital decaying NS-WD system, while those with $'$ are the corresponding quantities for the gravitational scalar radiation.

Then the Klein-Gordon equation of gravitons reads
\begin{equation}
[\Box-m_s^2\frac{\phi_0}{M_{pl}\Delta a}]\bar{h}_{\mu\nu} = 0.
\end{equation}
Consequently, we can find that the gravitons acquire a mass, which is expressed as
\begin{equation}
m_g^2 = m_s^2\frac{\phi_0}{M_{pl}\Delta a}.
\end{equation}
The asymptotic value of the gravitational scalar radiation field near spatial infinity $\phi_0$ is constrained to $\phi_0^{\it max} \lesssim 10^{-3}$ in weak-field tests, for a coupling strength of about $-6$ \cite{Damour:1996ke}. In the binary-pulsar measurements \cite{Damour:1998jk, 2012MNRAS.423.3328F, 2013Sci...340..448A}, $\phi_0$ is constrained to very close to zero, and it is usually to be taken as $\phi_0G^{1/2} = 10^{-3} - 10^{-5}$. Because of a less compactness and relatively lower surface gravity of WD, we consider $\phi_0G^{1/2} = 10^{-3}$ in NS-WD binary. Accordingly, we estimate the gravitons radiated from NS-WD binaries can acquire a mass of the order of $\sim 10^{-23} eV/c^2$. From Eq. (\ref{smass}), the mass of scalarons is a function of energy density of stars' scalar configurations and also depends on the strength of the scalar field and the scalar coupling strength. Therefore, the value of graviton mass $m_g$ in NS-WD systems mildly varies according to the microphysics of the interior of NS/WD and intrinsic properties of the binaries.

\section{Remarks and discussions}\label{sec:rd}

According to Einstein's general relativity, the gravitational waves propagate at speed of light. Therefore, the gravitons have zero mass. However, the recent events of GW 150914 detected by LIGO predicted an upper limit mass of $10^{-22} eV/c^2$ for graviton \cite{2016PhRvL.116f1102A}. Theoretically, the theories of massive gravity have been widely studied since Fierz-Pauli's massive spin-2 theory in 1939 \cite{1939HPA.12..3F, 1939RSPSA.173..211F}, which suffers from some problems, such as the Dam-Veltman-Zakharov discountinuity \cite{1970NuPhB..22..397V, 1970ZhPmR..12..447Z, 1970JETPL..12..312Z} and the existence of a ghostlike degree of freedom \cite{Boulware:1973my}. It has turned out that the theoretical problems involved for constructing a complete massive gravity theory are very subtle and challenging. A large of mountain efforts that dedicated to counting for the problems and the realization of a smooth theory of massive gravity \cite{Hinterbichler:2011tt, 2014LRR....17....7D}. The Higgs mechanism for gravitons was discussed in the scenario of spontaneous symmetry breaking of diffeomorphisms through the condensation of scalar fields \cite{Chamseddine:2010ub, Oda:2010gn}.

In NS-WD binary, a spontaneous scalarization, in analogy with the spontaneous magnetization in ferromagnet, tends to occur for the massive NS \cite{Damour:1993hw}, which changes the microphysics in the interior of NS and subsequently the dynamics of the system \cite{Damour:1996ke}. An inspiraling scalarized NS-WD binaries actually sources an emission of the dipolar gravitational scalar radiation \cite{Palenzuela:2013hsa}, which results in a gravitational scalar radiation field. It is the appearance of gravitational scalar radiation field and the induced gravitational scalar potential that plays the role of spontaneous breakdown of Lorentz invariance constructed in the framework of general relativity, which makes the NS-WD binary to be a natural laboratory to investigate the gravitational Higgs mechanism for associated particles \cite{Coates:2016ktu}. 

The scalar-tensor theory is then the alternative theory to general relativity that describes the dynamics of orbital decaying scalarized NS-WD binary. The effective gravitational scalar potential, consisting of a monotonically decreasing potential due to the self-interactions of gravitational scalar field and an exponential coupling between the scalar field and the NS/WD matter, contributes to an effective mass for the scalarons, which depends on the energy density of scalar configurations $\varepsilon_{\varphi_i}$. Consequently, the scalarized NS-WD system, encoded by a massive gravitational scalar field, is subject to a Yukawa type of potential described by Eq. (\ref{YukPot}), with an interaction range of $\lambda \sim \Delta a \sim 2R_{WD} \sim 10^6 m$. It is estimated that the mass of scalarons is of the order of $\sim 10^{-21} eV/c^2$. While the gravitational interactions encoded by the massive scalar mode is suppressed and screened in the thin-ring orbit of the binary by the screening parameter $\frac{\Delta a}{a}$. The gravitational scalar field directly interacts with the NS/WD matter via a conformal coupling $e^{-\alpha_i\phi/M_{pl}}$. The propagations of gravitational waves suffer from the condensation of massive gravitational scalar mode, with a Yukawa-corrected potential, in the 4-dimensional spacetime ($\vec{r}$ and $t$). Accordingly, the gravitons obtain a scalar-background-dependent mass term with value of the order of $10^{-23} eV/c^2$, which mildly varies with the microphysics in the interior of the stars and the intrinsic properties of the binaries. 

The NS-WD binaries are relatively wide systems, whose orbital binding energy may not be sufficient to activate dynamical scalarizations. However, a relatively more massive NS usually appears \cite{2011A&A...527A..83Z} in the system due to a recycling process \cite{2011A&A...526A..88W}. Therefore, the gravitational Higgs mechanism in NS-WD system actually results from the spontaneous scalarization of massive NS. As a consequence, the Lorentz invariance constructed in the framework of general relativity, with only quadruple radiations is spontaneously broken by an emission of dipolar gravitational scalar radiation. The spontaneous Lorentz symmetry breaking is responsible for the mass generation of associated articles, i.e. the radiated scalarons and gravitons in the process of orbital decay. The masses of scalarons and gravitons mildly vary with the environment due to the fluctuations of a compactness-dependent scalar energy density, which are thus different from the gravitational waves emitted during the last stages of inspiraling compact binary. As the magnitude of deviations from general relativity can depend non-linearly on the binding energy, the more massive NSs, e.g. PSRs J0348+0432 with mass of $1.97\pm 0.04 M_{\odot}$ and J1614-2230 with mass of $2.01\pm 0.04 M_{\odot}$, can be more promising systems used to probe the non-perturbative strong-field deviations away from general relativity, which is qualitatively very different compared to other binary-pulsar experiments. The effect is true even for double NS binaries that have small differences on their binding energies. 

A feature universal to all the forms of the massive graviton propagator is the effect of the mass on the dispersion relation, which makes the speed of the gravitational waves depend on the wave frequency $\omega_g$, i.e. $v(g)^2 = c^2(1 - \frac{m_g^2c^4}{\hbar^2\omega_g^2})$, and thus results in a time delay for the pulse signal of the radio pulsar in the system. We expect that the future experiments for electromagnetic and gravitational signal from NS-WD binaries performed by both pulsar timing and gravitational wave detectors can both test our results and provide precise gravitons mass estimates.

\begin{acknowledgments}
This work is supported by the Fundamental Research Funds
for the Central Universities (Grant no. 161gpy49) at Sun Yat-
Sen University and the Science and Technology Program of Guangzhou (20177100042050001).
\end{acknowledgments}


\vfill

\begin{thebibliography}{99}

\bibitem{2011A&A...527A..83Z} C.~M.~Zhang, J.~ Wang, Y.~H.~Zhao, H.~X.~Yin, L.~M.~Song, D.~P.~Menezes, D.~T.~Wickramasinghe, L.~Ferrario, P.~Chardonnet, Astron. Astrophys. 527, A83 (2011)

\bibitem{2010Natur.467.1081D} P.~B. Demorest, T. Pennucci, S.~M. Ransom, M.~S.~E. Roberts, J.~W.~T. Hessels, Nature, 467, 1081 (2010)

\bibitem{2013Sci...340..448A} J.~Antoniadis, P.~C.~C.~Freire, N.~ Wex, T.~M.~Tauris, R.~S.Lynch, M.~H.~van Kerkwijk,, M.~Kramer, C.~Bassa, V.~S.~Dhillon,  T.~Driebe, J.~W.~T.~Hessels, V.~M.~Kaspi, V.~I.~Kondratiev, N.~Langer, T.~R.~Marsh, M.~A.~McLaughlin, T.~T.~Pennucci, S.~M.~Ransom, I.~H.~Stairs, J.~van Leeuwen, J.~P.~W.~Verbiest, D.~G.~ Whelan, Science, 340, 448 (2013)

\bibitem{2011A&A...526A..88W} J. Wang, C. M. Zhang, Y. H. Zhao, Y. Kojima, H. X. Yin, L. M. Song, Astron. Astrophys., 526, A88 (2011)

\bibitem{Damour:1993hw}
  T.~Damour and G.~Esposito-Farese,
  Phys.\ Rev.\ Lett.\  {\bf 70}, 2220 (1993).

\bibitem{Damour:1996ke}
  T.~Damour and G.~Esposito-Farese,
  Phys.\ Rev.\ D {\bf 54}, 1474 (1996)

\bibitem{Palenzuela:2013hsa} 
  C.~Palenzuela, E.~Barausse, M.~Ponce and L.~Lehner,
  Phys.\ Rev.\ D {\bf 89}, no. 4, 044024 (2014)

\bibitem{Damour:1992we}
  T.~Damour and G.~Esposito-Farese,
  Class.\ Quant.\ Grav.\  {\bf 9}, 2093 (1992).
  
 \bibitem[Freire et al.(2012)]{2012MNRAS.423.3328F} P.~C.~C.~Freire, N.~ Wex, G.~Esposito-Far{\`e}se, J.~P.~W.~Verbiest, M.~Bailes, B.~A.~Jacoby, M.~Kramer, I.~H.Stairs, J.~Antoniadis, G.~H.~Janssen, Mon. Not. R. Astron. Soc., 423, 3328 (2012)
  
\bibitem{Khoury:2003rn} 
  J.~Khoury and A.~Weltman,
  Phys.\ Rev.\ D {\bf 69}, 044026 (2004)

\bibitem{Damour:1990tw}
  T.~Damour, G.~W.~Gibbons and C.~Gundlach,
  Phys.\ Rev.\ Lett.\  {\bf 64}, 123 (1990).
  
\bibitem{Salgado:1998sg} 
  M.~Salgado, D.~Sudarsky and U.~Nucamendi,
  Phys.\ Rev.\ D {\bf 58}, 124003 (1998)

\bibitem[Maggiore(2008)]{Maggiore2008} Michele Maggiore, "Gravitational Waves Volume 1: Theory and Experiments'' Oxford University Press, London, England (2008)

\bibitem{Wagoner:1970vr} 
  R.~V.~Wagoner,
  Phys.\ Rev.\ D {\bf 1}, 3209 (1970).

\bibitem[Will(1993)]{Will1993} C. M. Will, "Theory and Experiment in Gravitational Physics" Cambridge University Press, Cambridge, England (1993)

\bibitem{Alsing:2011er} 
  J.~Alsing, E.~Berti, C.~M.~Will and H.~Zaglauer,
  Phys.\ Rev.\ D {\bf 85}, 064041 (2012)

\bibitem{Damour:1998jk} 
  T.~Damour and G.~Esposito-Farese,
  Phys.\ Rev.\ D {\bf 58}, 042001 (1998)
  
 \bibitem[Abbott et al.(2016)]{2016PhRvL.116f1102A} Abbott, B.~P., Abbott, R., Abbott, T.~D., et al.  Phys. Rev. Lett., 116, 061102 (2016) 

\bibitem[Fierz \& Pauli(1939a)]{1939HPA.12..3F} Fierz, M., Helv. Phys. Acta, 12, 3 (1939)

\bibitem[Fierz \& Pauli(1939b)]{1939RSPSA.173..211F} Fierz, M., \& Pauli, W., Proceedings of the Royal Society of London Series A, 173, 211 (1939)

\bibitem[van Dam \& Veltman(1970)]{1970NuPhB..22..397V} van Dam, H., \& Veltman, M., Nucl. Phys. B, 22, 397 (1970)

\bibitem[Zakharov(1970a)]{1970ZhPmR..12..447Z} Zakharov, V.~I., ZhETF Pisma Redaktsiiu, 12, 447 (1970)

\bibitem[Zakharov(1970b)]{1970JETPL..12..312Z} Zakharov, V.~I., Soviet Journal of Experimental and Theoretical Physics Letters, 12, 312 (1970) 
  
\bibitem{Boulware:1973my} 
  D.~G.~Boulware and S.~Deser,
  Phys.\ Rev.\ D {\bf 6}, 3368 (1972).

\bibitem{Hinterbichler:2011tt} 
  K.~Hinterbichler,
  Rev.\ Mod.\ Phys.\  {\bf 84}, 671 (2012)

\bibitem[de Rham(2014)]{2014LRR....17....7D} C. de Rham, Living Reviews in Relativity, 17, 7 (2014)


\bibitem{Chamseddine:2010ub} 
  A.~H.~Chamseddine and V.~Mukhanov,
  JHEP {\bf 1008}, 011 (2010)
  doi:10.1007/JHEP08(2010)011
  [arXiv:1002.3877 [hep-th]].

\bibitem{Oda:2010gn} 
  I.~Oda,
  Phys.\ Lett.\ B {\bf 690}, 322 (2010)
  doi:10.1016/j.physletb.2010.05.048
  [arXiv:1004.3078 [hep-th]].

\bibitem{Coates:2016ktu} 
  A.~Coates, M.~W.~Horbartsch and T.~P.~Sotiriou,
  Phys.\ Rev.\ D {\bf 95}, no. 8, 084003 (2017)
  doi:10.1103/PhysRevD.95.084003
  [arXiv:1606.03981 [gr-qc]].

\end{thebibliography}
\end{document}